\documentclass{ws-procs961x669}            

\renewcommand*{\bi}{\begin{itemize}}
\newcommand*{\ei}{\end{itemize}}
\newcommand*{\be}{\begin{eqnarray}}
\newcommand*{\ee}{\end{eqnarray}}
\newcommand*{\bea}{\begin{eqnarray}}
\newcommand*{\eea}{\end{eqnarray}}

\begin{document}
\title{Aspects of Quark Orbital Angular Momentum}

\author{M. Burkardt}

\address{Physics Department, New Mexico State University,\\
Las Cruces, NM 88003, U.S.A.\\
$^*$E-mail: burkardt@nmsu.edu}

\begin{abstract}
The difference between the quark orbital angular momentum (OAM) defined in light-cone gauge (Jaffe-Manohar) compared to
defined using a local manifestly gauge invariant operator (Ji) is interpreted in terms of the change in quark OAM as the quark leaves the 
target in a DIS experiment. We also discuss the possibility to measure quark OAM directly using twist 3 GPDs, and to calculate quark OAM in lattice QCD.

\end{abstract}


\bodymatter
\section{Introduction}
GPDs provide information about the longitudinal momentum and the transverse
position of partons\cite{mb1}. This is reminiscent or orbital angular momentum (OAM), which also requires momentum and position in orthogonal directions. It is thus not completely surprising that GPDs can be related to angular momentum as\cite{JiPRL}
\be
J_q=\frac{1}{2}\int_0^1 dx\, x\left[H^q(x,0,0)+E^q(x,0,0)\right].
\ee
Here $J_q=L_q+\frac{1}{2}\Delta q$ where $\frac{1}{2}\Delta q$ is the quark spin contribution and
\begin{equation}
{ L}_q^z= \int d^3r \langle PS| q^\dagger \left({\vec r} \times \frac{1}{i}{\vec D}
\right)^zq |PS\rangle /\langle PS|PS\rangle
\label{M012}
\end{equation}
in a nucleon state polarized in the $+\hat{z}$ direction. Here
${\vec D}={\vec \partial}-ig{\vec A}$ is the gauge-covariant
derivative.

\section{Angular Momentum Decompositions}
Unfortunately, there are complications due to issues related to the definition of OAM.
To illustrate this point, let us start from a definition of photon total angular momentum based on the Poynting vector. Performing some integration by parts and making use of $\vec \nabla \cdot \vec E =e\psi^\dagger\psi$ one can
rewrite $\vec J_\gamma$ as a term that can be interpreted as photon OAM, a term that cancels the contribution involving
$\vec A$ in $\vec L_q$, and a term that can be interpreted as photon spin:
\begin{eqnarray}\label{eq:Jg}
{\vec J}_\gamma &=& \int d^3r \,{\vec r} \times \left({\vec E}\times{\vec B}
\right) =  \int d^3r \,\left[E^j \left({\vec r}\times {\vec \nabla}
\right) A^j
+ \left({\vec r}\times {\vec A}\right) e\psi^\dagger \psi
+ {\vec E}\times {\vec A}\right]
\end{eqnarray}
Eq. (\ref{eq:Jg}) illustrates that decomposing ${\vec J}_\gamma$ into spin and orbital also
shuffles angular momentum from photons to electrons!

Jaffe and Manohar have proposed an alternative decomposition of the
nucleon spin, which does have a partonic interpretation, and in which also two terms, 
$\frac{1}{2}\Delta q$ and $\Delta G$,
are experimentally accessible \cite{JM}
\begin{equation}
\frac{1}{2}=\frac{1}{2}\sum_q\Delta q + \sum_q {\cal L}^q+
\Delta G + {\cal L}^g.
\label{eq:JJM}
\end{equation}
In this decomposition the quark OAM is defined as 
\begin{equation}
{\cal L}^q \equiv \int d^3r \langle PS|q^\dagger_+\!\left({\vec r}\times \frac{1}{i}{\vec \partial}
\right)^z \!\!q_+  |PS\rangle / \langle PS|PS\rangle,
\label{M+12}
\end{equation}
where light-cone gauge $A^+=0$ is implied. Although Eq. (\ref{M+12}) is not manifestly gauge invariant as written, gauge invariant extensions can be defined
\cite{lorce,hatta}. Indeed, manifestly gauge invariant definitions for each of the terms
in Eq. (\ref{eq:JJM}) exist,  which, with the exception of $\Delta q$, involve matrix elements of
nonlocal operators. In light-cone gauge those nonlocal operators reduce to a local 
operator, such as Eq. (\ref{M+12}).

\section{OAM from Wigner Distributions}

Given the fact that different spin decompositions and corresponding definitions for quark OAM are possible raises the question about the physical interpretation of those differences. In this effort significant progress has been made based on 5-dimensional Wigner distributions.

Wigner distributions can be defined as 
off forward matrix elements of non-local
correlation functions\cite{jifeng,Metz,lorce} with $P^+=P^{+\prime}$, $P_\perp = -P_\perp^\prime = \frac{q_\perp}{2}$
\begin{eqnarray}\label{eq:wigner}
\!\!\!\!\!\!\!W^{\cal U}\!(x,\!{\vec b}_\perp,\! {\vec k}_\perp)\!
\equiv \!\!\!
\int \!\!\frac{d^2{\vec q}_\perp}{(2\pi)^2}\!\!\int \!\!\frac{d^2\xi_\perp d\xi^-\!\!\!\!}{(2\pi)^3}
e^{-i{\vec q}_\perp \!\!\cdot {\vec b}_\perp}\!
e^{i(xP^+\xi^-\!\!-{\vec k}_\perp\!\!\cdot{\vec \xi}_\perp)}
\langle P^\prime S^\prime |
\bar{q}(0)\Gamma {\cal U}_{0\xi}q(\xi)|PS\rangle .
\end{eqnarray}
Throughout this paper, we will chose ${\vec S}={\vec S}^\prime = \hat{\vec z}$. Furthermore, we will focus on the 'good' component by selecting $\Gamma=\gamma^+$.
To ensure manifest gauge invariance, a Wilson line gauge link 
${\cal U}_{0\xi}$ connecting the quark field operators at position $0$ and $\xi$ is included. The issue of choice of path
for the Wilson line will be addressed below. 

In terms  of  Wigner distributions,  TMDs and OAM can be defined 
as \cite{lorce}
\begin{eqnarray}
f(x,{\vec k}_\perp) &=& \int dx d^2{\vec b}_\perp d^2{\vec k}_\perp {\vec k}_\perp 
W^{\cal U}(x,{\vec b}_\perp,{\vec k}_\perp)\\
L_{\cal U}&=& \int dx d^2{\vec b}_\perp d^2{\vec k}_\perp \left({\vec b}_\perp \times {\vec k}_\perp \right)^z
W^{\cal U}(x,{\vec b}_\perp,{\vec k}_\perp).
\nonumber
\end{eqnarray}
No issues with Heisenberg's uncertainty principle arise here: only perpendicular combinations of position ${\vec b}_\perp$ and momentum ${\vec k}_\perp$ are
needed simultaneously in order to evaluate the integral for
$L_{\cal U}$.

A straight line connecting $0$ and $\xi$ for the Wilson line in ${\cal U}_{0\xi}$ results in
\cite{jifeng}
\begin{eqnarray}
L^q_{straight}
&=&
L^q_{Ji}.
\label{eq:LJi}
\end{eqnarray}
However, depending on the context, other choices for the path in the Wilson link ${\cal U}$ should be made. Indeed for TMDs probed in SIDIS the path should be taken to be a straight line to $x^-=\infty$
along (or, for regularization purposes, very close to) the light-cone. This particular choice ensures proper inclusion of the FSI experienced by the struck quark as it leaves the nucleon
along a nearly light-like trajectory in the Bjorken limit. However, a Wilson line to
$\xi^-=\infty$, for fixed ${\vec \xi}_\perp$ is not yet sufficient to render Wigner distributions
manifestly gauge invariant, but a link at $\xi^-=\infty$ must be included to ensure manifest
gauge invariance. While the latter may be unimportant in some gauges, it is crucial in
light-cone gauge for the description of TMDs relevant for SIDIS. 

\begin{figure}
\includegraphics[scale=0.5]{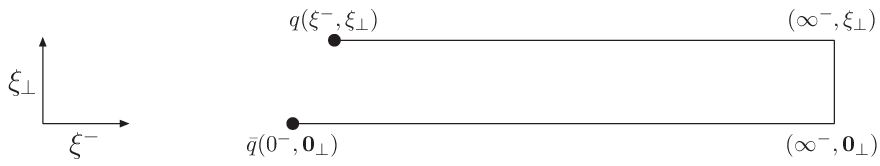}
\caption{Illustration of the path for the Wilson line gauge link ${\cal U}^{+LC}_{0\xi}$ entering  $W^{+LC}$ }
\label{fig:staple}
\end{figure}

Let ${\cal U}^{+LC}_{0\xi}$ be the Wilson path ordered exponential obtained by first taking
a Wilson line from $(0^-,{\vec 0}_\perp)$ to $(\infty,{\vec 0}_\perp)$, 
then to $(\infty,{\vec \xi}_\perp)$, and then to $(\xi^-,{\vec \xi}_\perp)$, with each segment being a straight line (Fig. \ref{fig:staple}) \cite{hatta}. 
The shape of the segment at $\infty$ is irrelevant as the gauge field is pure gauge there, but it is still necessary to include a connection at $\infty$ and for
simplicity we pick a straight line. 
In light-cone gauge $A^+=0$, only the segment at $\xi^-=\pm \infty$ contributes and 
the OAM looks similar to the local manifestly gauge invariant expression, except
\be
{\vec r}\times {\vec A}({\vec r}) \longrightarrow {\vec r}\times {\vec A}(r^-=\pm \infty, {\bf r}_\perp).
\ee

This observation is crucial for understanding the difference 
between the Ji vs. Jaffe-Manohar OAM, which in light-cone gauge\footnote{As $L^q$ involves a manifestly gauge invariant local operator, it can be evaluated in any gauge.}
involves only the replacement ${ A}_\perp^i({\vec r}) \longrightarrow {A}_\perp^i(r^-=\pm \infty, {\bf r}_\perp)$.
Using
\begin{eqnarray}
{A}^i_\perp (r^-\!\!=\infty,{\bf r}_\perp)-{ A}^i_\perp (r^-,{\bf r}_\perp)
=\!\!\int_{r^-}^\infty\!\!\!\!\! dz^-
\partial_- {A}^i_\perp (z^-,{\vec r}_\perp)
= \!\!\int_{r^-}^\infty\!\!\!\!\! dz^- G^{+i}(z^-,{\vec r}_\perp)
\label{eq:kp}
\end{eqnarray}
where $G^{+\perp}=\partial_-A^\perp$ is the gluon field strength tensor in $A^+=0$ gauge. 
\be
{\cal L}^q-L^q = -g \!\!\int \!\!d^3x\!\left\langle P\!,\!S\right|\!
\bar{q}({\vec x})\!\gamma^+\!\!
\left[{ {\vec x}\! \times\! \!
\int_{x^-}^\infty \!\!\!\!\!dr^- F^{+\perp}(r^-,{\bf x}_\perp)
}\right]^z\!\!\!\!
q({\vec x}) \!\left| P\!,\!S\right\rangle/ \langle PS|PS\rangle
\label{eq:torque}
\ee
Note that 
\begin{equation}
-\sqrt{2}gG^{+y}\equiv -gG^{0y}-gG^{zy} = g\left(E^y-B^x
\right)
=g\left({\vec E}+{\vec v}\times {\vec B}\right)^y
\end{equation}
yields the $\hat{y}$ component of the color Lorentz force acting on a particle that moves with the velocity of light in the $-\hat{z}$ direction (${\vec v}=(0,0,-1)$) --- which is the direction of the 
momentum transfer in DIS. Thus the difference between the Jaffe-Manohar and Ji OAMs
has the semiclassical interpretation of the change in OAM due to the torque from the FSI as the quark leaves the target:\cite{mb:torque}
while $L^q$ represents the local and manifestly gauge invariant OAM of the
quark {\it before} it has been struck by the $\gamma^*$, ${\cal L}^q$ represents 
the gauge invariant OAM {\it after} it has left the nucleon and moved to $r^-=\infty$.

To convince oneself that the net effect from such a torque can be nonzero, one can consider an ensemble of quarks ejected from 
a nucleon with a color-magnetic field aligned with its spin.

\section{Torque in Spectator Models}
Given the fact that the Jaffe-Manohar and Ji definitions of quark orbital angular momentum are differ by the potential angular momentum raises the question as how significant that difference actually is.
In Ref. \cite{JiElectron} it was shown that to ${\cal O}(\alpha)$ in QED
$L_{Ji}={\cal L}_{JM}$. This corrects an earlier result \cite{BC} where the contribution from states with longitudinally polarized Pauli-Villars photons had been omitted.

In order to assess the significance of FSI effects for quark OAM we thus considered the effects of the vector potential in the scalar diquark model. While we do not consider this model a good approximation for QCD, it has been very useful in several respects:
Most importantly, the model allows for a fully Lorentz invariant calculation of 'nucleon' matrix elements --- which is not the case for almost all other models for nucleon structure.
Furthermore, this was the first model that clearly illustrated the role of FSI and the Sivers effect\cite{Sivers} in SIDIS and DY \cite{BHS}. 

We found that a nonzero potential angular momentum arises at the same order
as transverse single-spin asymmetries \cite{BHS}, i.e. one photon/gluon exchange.\cite{mb:torque}

{\bf Acknowledgements:}
This work was supported by the DOE under grant number 
DE-FG03-95ER40965


\begin{thebibliography}{10}

\bibitem{mb1}	
  M.~Burkardt,
  Phys.\ Rev.\ D {\bf 62}, 071503 (2000)
  Erratum: [Phys.\ Rev.\ D {\bf 66}, 119903 (2002)]; 
  M.~Burkardt,
  Int.\ J.\ Mod.\ Phys.\ A {\bf 18}, 173 (2003); 
  M.~Diehl,
  Phys.\ Rept.\  {\bf 388}, 41 (2003).
\bibitem{JiPRL} X. Ji, 
{Phys. Rev. Lett.} {\bf 78} (1997) 610.
\bibitem{JM} R.L. Jaffe and A. Manohar, 
{ Nucl. Phys. B} {\bf 337} (1990) 509.
\bibitem{jifeng} A.V. Belitsky, X. Ji, and F. Yuan, 
{ Phys. Rev. D} {\bf 69} (2004) 074014.
\bibitem{lorce} C. Lorc\'e and B. Pasquini, 
{ Phys. Rev. D} {\bf 84} (2011) 014015.
\bibitem{hatta} Y. Hatta,
{Phys. Rev. D} {\bf 84}, 041701 (2011); 
{ Phys. Lett. B} {\bf 708} (2012) 186.
\bibitem{mb:torque} M. Burkardt, 
{ Phys. Rev. D} {\bf 88} (2013) 1.
\bibitem{JiElectron} X. Ji {et al.}, 
Phys.Rev. D93 (2016) 054013.
\bibitem{BC} M. Burkardt and H. BC, 
{ Phys. Rev. D} {\bf 79} (2009) 071501.


\bibitem{Sivers} 
  D.~W.~Sivers,
  Phys.\ Rev.\ D {\bf 43}, 261 (1991).
\bibitem{BHS} S.J. Brodsky, D.-S. Hwang, and I. Schmidt, 
Phys.Lett. B530 (2002) 99.
\bibitem{mb:torque}
  D.~A.~Amor-Quiroz, M.~Burkardt and C.~Lorcé,
  PoS DIS {\bf 2019}, 168 (2019)




 




  



\end{thebibliography}
\end{document}